\documentclass[final,5p,times,twocolumn]{elsarticle}

 \usepackage{graphicx}

\usepackage{amssymb}

\journal{Physica E}

\begin{document}

\begin{frontmatter}



\title{Landau level broadening in graphene with long-range disorder\\
--- Robustness of the $n=0$ level}


\author[toho]{T. Kawarabayashi}
\author[tsukuba]{Y. Hatsugai}
\author[tokyo]{H. Aoki}

\address[toho]{Department of Physics, Toho University, Funabashi 274-8510, Japan}
\address[tsukuba]{Institute of Physics, University of Tsukuba, Tsukuba 305-8571, Japan}
\address[tokyo]{Department of Physics, University of Tokyo, Tokyo 113-0033, Japan}

\begin{abstract} 
Broadening of the Landau levels in graphene and the associated 
quantum Hall plateau-to-plateau transition are investigated numerically.
For correlated bond disorder, the graphene-specific 
$n=0$ Landau level of the Dirac fermions becomes anomalously sharp 
accompanied by the Hall transition 
exhibiting a fixed-point-like criticality. 
Similarly anomalous behavior for the $n=0$ Landau level 
is also shown to occur in correlated random 
magnetic fields, which suggests that 
the anomaly is generic to disorders that preserve the chiral symmetry.
\end{abstract}


\end{frontmatter}


\section{Introduction}
After the seminal 
discovery of the characteristic quantum Hall effect (QHE) in graphene \cite{Geim,Kim}, 
attention has been focused on 
the effect of disorder \cite{KA,SM,OGM,NRKMF,NGPNG}. 
Randomness, which is important in the QHE in ordinary 
2DEG, is particularly crucial since it 
affects such key factors in the lattice (as opposed 
to Dirac field) models of graphene 
such as the {\it chiral} (A-B sub-lattice) symmetry\cite{HFA}, 
and the scattering between the valleys (K and K').  
Note that the valley degeneracy has a topological origin as
the doubling of the massless Dirac fermions on the honeycomb lattice.
It is then natural that the nature of disorder crucially affects the electronic structure.
A potential disorder (including random site energies) 
destroys the chiral symmetry, whereas a disorder 
in bonds respects the symmetry, 
while the spatial correlation in the disorder 
controls the inter-valley scattering. 
In order to explore these, here 
we adopt the honeycomb lattice rather than the effective Dirac model 
to investigate how the spatial correlation of disorder affects 
the Landau level structure, especially the stability of the 
$n=0$ Landau level which is essential to the characteristic 
QHE in graphene. 

In the case of a potential disorder 
with degraded chiral symmetry, 
the quantum 
Hall transition at the $n=0$ Landau level has been shown to be robust 
in an effective Dirac model \cite{NRKMF}. 
In such a model, however, the criticality of the transition at $n=0$ is described by the ordinary quantum Hall transition, with nothing special about 
the criticality at the $n=0$ transition. 
In the previous paper \cite{KHA}, we have shown that for the bond disorder, which preserves the chiral symmetry, the criticality 
at $n=0$ transition is  anomalously sensitive to the spatial correlation of disorder. As soon as the correlation length 
of the bond disorder exceeds few lattice constants of the honeycomb 
lattice, the $n=0$ Landau level becomes anomalously 
sharp and the associated Hall transition exhibits the anomalous criticality, 
which 
corresponds to a fixed point of the 
Dirac fermion with chiral symmetry \cite{LFSG}.
In the present paper, we present 
another example to reinforce our arguments that 
the anomalous Hall transition is a general property of 
the chiral-symmetric random systems.  
Namely, we consider systems with random phases in the transfer integrals, which  
corresponding to the case of 
random magnetic fields piercing the hexagons in honeycomb lattice.

\section{Models}
The tight-binding model for the honeycomb lattice is described by the Hamiltonian 
$H=\sum_{\langle i,j\rangle}t_{ij}{\rm e}^{-2\pi{\rm i}\theta_{ij}} c_i^{\dagger}c_j + {\rm h.c.}$, where the transfer integral $t_{i,j}$ is real while
the Peierls phases $\theta_{ij}$ satisfy the requirement that the sum of the phases around a hexagon is 
equal to the magnetic flux piercing a hexagon in units of the flux quantum $\phi_0= h/e$. 

Random bonds are introduced as $t_{ij} = t + \delta t_{ij}$, where $t$ and $\delta t_{ij}$ are the 
uniform and the random components, respectively. 
The random components $\delta t_{ij}$ are assumed to have a Gaussian distribution, 
$P(\delta t) = {\rm e}^{-\delta t^2/2\sigma_t^2}/\sqrt{2\pi\sigma_t^2}$, with a variance $\sigma_t$ and a spatial correlation,
$$
 \langle \delta t_{ij} \delta t_{kl}
 \rangle = \langle \delta t^2 \rangle \exp(-|\mbox{\boldmath $r$}_{ij} -
 \mbox{\boldmath $r$}_{kl}|^2/4\eta_t^2),
$$
with a correlation length $\eta_t$, where $\mbox{\boldmath $r$}_{ij}$ denotes the 
position  of the bond $t_{ij}$ and $\langle \rangle$ the ensemble average. 

On the other hand, random magnetic fields are introduced as 
$\phi (\mbox{\boldmath $r$}) = 
\phi + \delta \phi(\mbox{\boldmath $r$})$.   
This type of disorder, being another disorder in bonds, also 
preserves the chiral symmetry.  Here $\phi$ represents 
the uniform part of the magnetic field, while $\delta \phi(\mbox{\boldmath $r$})$ the 
random magnetic fluxes for hexagons each 
located at position $\mbox{\boldmath $r$}$.
We assume that the random component $\delta \phi$ obeys a 
Gaussian distribution with a variance $\sigma_{\phi}$ and a spatial correlation 
$$
 \langle \delta \phi(\mbox{\boldmath $r$}_i)  \delta \phi(\mbox{\boldmath $r$}_j) 
 \rangle = \langle \delta \phi^2 \rangle \exp(-|\mbox{\boldmath $r$}_{i} -
 \mbox{\boldmath $r$}_{j}|^2/4\eta_{\phi}^2)
$$
with a correlation length $\eta_{\phi}$\cite{KOOKSK}.  In the following, all lengths are measured in units of the bond length $a$ 
of the honeycomb lattice.

Let us recapitulate how these types of disorder preserve 
the chiral symmetry (even for each realization of disorder).  	
There exists a local unitary operator $\gamma$ that anti-commutes with the Hamiltonian, $\{ \gamma, H\}=0$, which defines the 
chiral symmetry, and $\gamma ^2 = 1$.
In the case of the honeycomb lattice, the lattice sites can be decomposed into $A$ and $B$ sublattices .  The operator 
$\gamma$ can then be given as  $\gamma=\exp({\rm i}\pi \sum_{i\in A}c_i^{\dagger}c_i )$, where $\sum_{i\in A}$ denotes
the summation over the A sublattice sites. With this choice of $\gamma$ 
we can readily verify that the fermion operator $c_i$
is transformed as $\gamma c_i \gamma^{-1} = -c_i$ for $i\in A$ and $\gamma c_i \gamma^{-1} = c_i$ for $i\in B$.  Due to this chiral symmetry, the energy levels always appear in pairs $\{ E, -E\}$ even in the presence of disorder. 
For instance, if we have an energy eigenstate 
$\psi_E$ satisfying $H\psi_E = E\psi_E$, the state $\gamma \psi_E$ is an eigenstate with energy $-E$ since $H\gamma \psi_E = 
-\gamma H\psi_E = -E \gamma \psi_E$.  
A special situation arises for zero-energy states, for which the eigenstates $\psi_{E=0}$ and $\gamma \psi_{E=0}$ are degenerated.
The eigenstates can then be made simultaneous eigenstates 
$\psi_{\pm} = (1\pm\gamma)\psi_{E=0}$ of the operator $\gamma$ 
with $\gamma \psi_{\pm} = \pm \psi_{\pm})$.  
The eigenstate $\psi_-(\psi_+)$ has non-zero amplitudes 
only on the $A(B)$ 
sublattice. The fact that the zero-energy states can be an eigenstate of 
$\gamma$ implies that 
the zero-energy states has an extra symmetry, so that 
the criticality at zero energy can 
be especially sensitive to the presence or otherwise 
of the chiral symmetry \cite{AZ,EM}.
Obviously, the potential disorder breaks this symmetry.

The density of states 
$\langle \rho_i \rangle = -\sum_i{\rm Im} G_{ii}(E+{\rm i}\epsilon)/N\pi$ is evaluated by the Green function \cite{SKM} 
$G_{ii}(E+{\rm i}\epsilon) = \langle i | (E-H+{\rm i}\epsilon)^{-1}|i\rangle$, where $N$ stands for the total number of sites and 
$\epsilon$ an infinitesimal imaginary part of energy for evaluating the Green function numerically. We have carried out calculations 
for values of $\epsilon /t$ reduced from $0.01$ down 
to $6.25\times 10^{-4}$ to confirm that the anomaly in 
the density of states at 
the $n=0$ Landau level  discussed below is not affected by the value of 
$\epsilon$. The system size considered is a 
$L_x \times L_y$ rectangular system with periodic boundaries in $y$ direction, where $x$ axis is assumed to be 
parallel to the zigzag direction of the honeycomb lattice.
In actual calculations, the Landau gauge for the corresponding bricklayer lattice \cite{WFAS} is adopted.

\section{Numerical Results --- Random Bonds}

\begin{figure}
\includegraphics[scale=0.3]{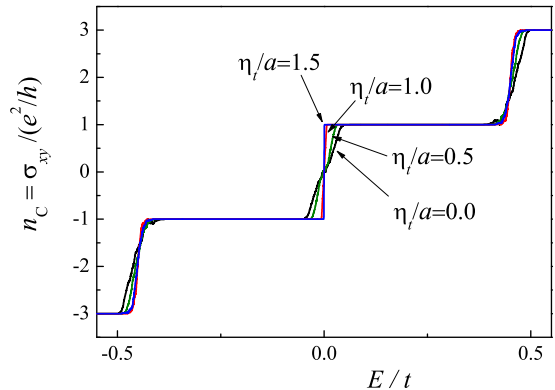}
\caption{The averaged Chern number $n_C$ 
(the Hall conductivity ($\sigma_{xy}/(e^2/h)$)) as a function of the Fermi energy ($E/t$). 
The Chern number for each realization of disorder is evaluated by using the lattice gauge technique\cite{KHA,Hatsugai1,Hatsugai2,FHS}.  
The results  for the correlation length $\eta_t/a=0$, $0.5$, $1.0$, and $1.5$ are shown 
for plateaus $n_C=\sigma_{xy}/(e^2/h)=-3$, $-1$, $1$, 
$3$ with the disorder strength 
$\sigma_t /t =0.12$, the magnetic field $\phi/\phi_0=1/50$, and an average over 300 samples for 
the system size $L_x/(\sqrt{3}a/2) = L_y/(3a/2)=20$.  
}
\label{fig0}
\end{figure}

\begin{figure}
\includegraphics[scale=0.35]{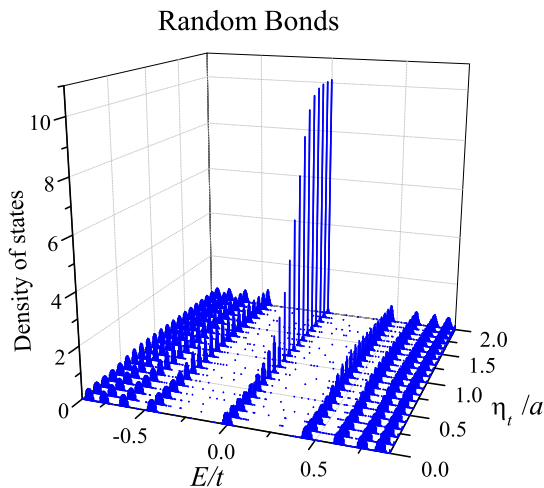}
\caption{The density of states for various values of 
the correlation length $\eta_t/a$ of the random bonds. The system-size is 
$L_x/(\sqrt{3}a/2) = 5000$, $L_y/(3a/2)=100$. The imaginary part of energy $\epsilon /t=6.25 \times 10^{-4}$.}
\label{fig1}
\end{figure}

First, we consider the random bonds in a uniform magnetic field ($\sigma_{\phi}=0$). In the previous paper \cite{KHA}, we 
have clearly demonstrated for such a model that when the correlation length $\eta_t$ exceeds few bond lengths, the 
$n=0$ Landau level becomes anomalously sharp and the associated quantum Hall transition shows an almost 
exact fixed-point-like criticality even in a finite system (Fig. \ref{fig0}).  
This sharply contrasts with the result for the uncorrelated case $\eta_t=0$ , 
for which the  
$n=0$ Landau level is broadened in the same way as $n\neq 0$ Landau levels.  The anomalous sensitivity to the 
correlation of disorder occurs only for the $n=0$ Landau level. 
We have examined the density of states for other values of $\sigma_t$ and $\phi$, and confirmed that the anomaly at the $n=0$ Landau 
level is commonly observed for $\eta_t/a \geq 1$. Here we show the density of states as a function of the Fermi energy and the 
correlation length $\eta_t$ for $\sigma_t/t=0.058$ and $\phi/\phi_0=1/50$
in Fig. \ref{fig1}. 
It is again 
clearly seen, as has been seen in the case of $\sigma_t /t = 0.12$ \cite{KHA}, 
that the $n=0$ Landau level becomes anomalously sharp compared to other Landau levels as soon as the 
correlation length $\eta_t$ is greater than the lattice constant.

In actual graphene samples, the scale of ripples is estimated to be of order of $10$ nanometers \cite{Meyer,Geringer}, which is 
much greater than the lattice constant $a \sim 1.42 \AA$ \cite{NGPNG}, and the magnitude of disorder is likely to be much smaller.
Our result therefore clearly indicates that the bond disorder induced by ripples should not broaden the $n=0$ Landau level.

\section{Numerical Results --- Random Magnetic Fields}

\begin{figure}
\includegraphics[scale=0.35]{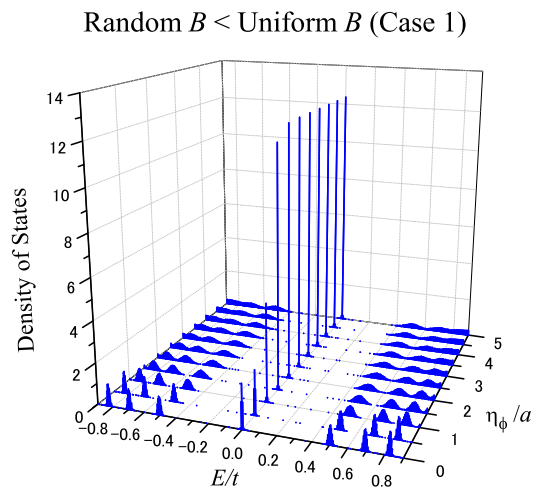}
\caption{The density of states is plotted for various values of 
the correlation length $\eta_{\phi}/a$ when the fluctuation of the magnetic field is smaller than its mean ($\sigma_{\phi} < \phi$). 
The parameters are $\sigma_{\phi}/\phi_0= 0.0058 < \phi/\phi_0=1/41 \simeq 0.024$ for a system size $L_x/(\sqrt{3}a/2) = 5000$, $L_y/(3a/2)=82$ and the imaginary part of energy $\epsilon /t=6.25 \times 10^{-4}$.}
\label{fig2}
\end{figure}

Let us next show the results for another model, namely the random magnetic field model with $\sigma_{\phi} \neq 0$, where no randomness 
is assumed for the amplitude of the 
transfer energies ($\sigma_t=0$).
For this model, we consider both the small random field case, $\sigma_{\phi} < \phi$ (case 1), and the large random field case, $\sigma_{\phi} > \phi$ (case 2). 
For $ \sigma_{\phi} < \phi$ the distinct Landau level structure is present 
(Fig. \ref{fig2}).  The $n=0$ Landau level is again anomalously sensitive to the spatial 
correlation of random magnetic fields: 
When the correlation length $\eta_{\phi}$ 
exceeds few lattice constants, the width of the $n=0$ Landau level becomes a sharp, delta-function-like peak.  
Indeed, the shape of the $n=0$ Landau level for $\eta_{\phi}/a >2.5$ is 
almost exactly given by the Lorentzian distribution $(\phi/\phi_0)(\epsilon/\pi)/(E^2+\epsilon^2)$, indicative of 
a  zero intrinsic width.  This again contrasts with the $n\neq 0$ Landau 
level.  

\begin{figure}
\includegraphics[scale=0.35]{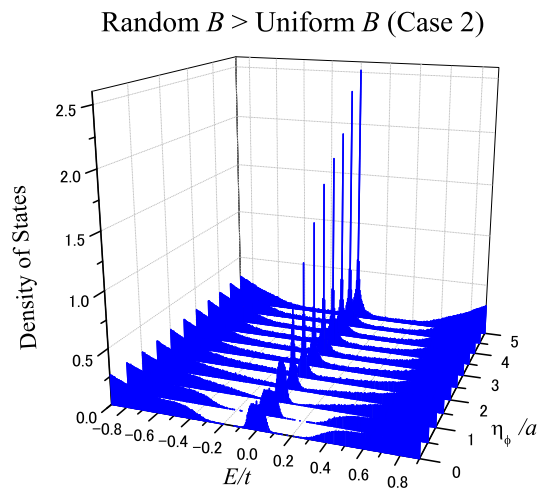}
\caption{As above when the fluctuation of the magnetic field is greater than its mean ($\sigma_{\phi} > \phi$). 
The parameters are $\sigma_{\phi}/\phi_0= 0.058 > \phi/\phi_0=1/41 
\simeq 0.024$ ($\phi < \sigma_{\phi}$) for a 
system size $L_x/(\sqrt{3}a/2) = 5000$, $L_y/(3a/2)=82$ and 
the imaginary part of energy $\epsilon/t=6.25 \times 10^{-4}$.}
\label{fig3}
\end{figure}

For $ \sigma_{\phi} > \phi$ (case 2), on the other hand, the Landau level structure is mostly washed out for $n\neq 0$ Landau levels, which is 
not surprising since the fluctuation of the magnetic field is larger than its mean value.  Surprisingly, $n=0$ Landau level exhibits an anomalous 
behavior, where the delta-function-like behavior arises when 
the correlation length of the random magnetic 
field exceed few lattice constants, 
even though the field fluctuation width is much greater than 
the average field (Fig.\ref{fig3}).

\section{Discussions and Conclusions}

\begin{figure}
\includegraphics[scale=0.35]{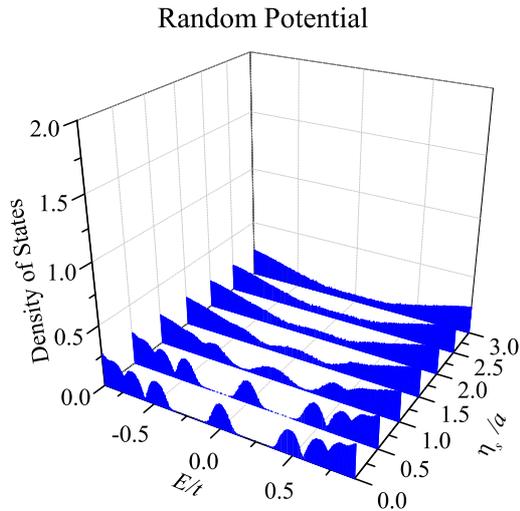}
\caption{The density of states in a random potential 
for various values of the correlation length $\eta_s/a$.
The parameters are $\sigma_s/t = 0.29$, $\phi/\phi_0=1/41$ and $\sigma_t = \sigma_{\phi}=0$ for a 
system size $L_x/(\sqrt{3}a/2) = 5000$, $L_y/(3a/2)=82$ 
and the imaginary part of energy $\epsilon/t=6.25 \times 10^{-4}$.}
\label{fig4}
\end{figure}

In order to check that the above anomalies are 
a consequence of the preserved chiral symmetry, 
we have also evaluated the density of states for the case of the spatially correlated potential disorder, where the Hamiltonian is given 
by $H= \sum_{\langle ij\rangle}(t {\rm e}^{-2\pi\theta_{ij}}c_i^{\dagger}c_j + {\rm h.c.}) + \sum_i \varepsilon_i c_i^{\dagger}c_i$, where 
random site energies $\varepsilon_i$ are assumed to obey a Gaussian distribution with a variance $\sigma_s$ and a spatial correlation
$\langle \varepsilon_i \varepsilon_j\rangle = \sigma_s^2 \exp(-(\mbox{\boldmath $r$}_i - \mbox{\boldmath $r$}_j)^2/4\eta_s^2)$.
The result for $\sigma_s/t =0.29$ and $\phi/\phi_0=1/41$ in Fig. \ref{fig4} 
confirms that no anomaly exists at the 
$n=0$ Landau level.   As we increase 
the correlation length $\eta_s$ keeping the magnitude of disorder $\sigma_s$ fixed, all the Landau levels are broadened \cite{AU}.

To summarize, we have investigated the anomalous behavior of  the $n=0$ Landau level of the 
QHE system on a honeycomb lattice. Two types of bond disorder 
that preserve the chiral symmetry 
have been considered, namely 
the randomness in the magnitude of the transfer integrals (random bonds) and that in their phases (random magnetic fields). 
It is clearly demonstrated that in both cases 
the $n=0$ Landau level is anomalously sensitive to the spatial correlation of disorder, where an anomaly at the $n=0$ Landau level 
appears as soon as the correlation length exceeds few 
bond lengths. This indicates that the absence of the mixing between two valleys are essential to the 
anomaly at the $n=0$ Landau level. We have also confirmed that the
anomaly does not exist for the correlated potential disorder, where the chiral symmetry is broken. 
The results suggest that the 
anomaly at the $n=0$ Landau level is generic to the graphene system with long-range disorders that preserve the chiral symmetry.  
This implies that the bond disorder induced by ripples in graphene 
should not broaden the $n=0$ Landau level; conversely, 
if a broadening is observed in experiments, that must be caused by other origins, such as the potential disorder from 
charged impurities in substrates. Experimentally,
a narrower $n=0$ Landau level of grephene has been reported by measuring the activation energy gaps \cite{GZKPMM}, 
which may be 
related to the present analysis.
More elaborate analysis on the nature of the disorder is an 
interesting future problem.

We would like to thank Yoshiyuki Ono, Tomi Ohtsuki and Allan MacDonald 
for valuable discussions and comments. 
The work was supported in part by Grants-in-Aid for Scientific Research,
Nos. 20340098 (YH and HA)
 and 20654034 
 from JSPS and
Nos. 220029004 
and 20046002 
on Priority Areas from MEXT for YH. 



\end{document}